\documentclass{nature}

\usepackage{graphicx}
\usepackage{bm}% bold math
\usepackage{color}

\title{Direct observation of lattice symmetry breaking at the hidden-order transition in URu$_2$Si$_2$}

\author{S.~Tonegawa$^{1}$, S.~Kasahara$^{1}$, T.~Fukuda$^{2,3}$, K.~Sugimoto$^{4,5}$, N.~Yasuda$^4$, Y.~Tsuruhara$^1$, D.~Watanabe$^1$, Y.~Mizukami$^{1,6}$, Y.~Haga$^7$, T.\,D.~Matsuda$^{7,8}$, E.~Yamamoto$^7$, Y.~Onuki$^{7,9}$, H.~Ikeda$^{1,10}$, Y.~Matsuda$^1$ \& T.~Shibauchi$^{1,6,\dag}$}

\begin{document}
\maketitle
\begin{affiliations}
 \item Department of Physics, Kyoto University, Sakyo-ku, Kyoto 606-8502, Japan
 \item Quantum Beam Science Directorate, JAEA SPring-8, Sayo, Hyogo 679-5148, Japan
 \item Materials Dynamics Laboratory, RIKEN SPring-8, Sayo, Hyogo 679-5148, Japan
 \item Research \& Utilization Division, JASRI SPring-8, Sayo, Hyogo 679-5198, Japan
 \item Structural Materials Science Laboratory, RIKEN SPring-8, Sayo, Hyogo 679-5148, Japan
 \item Department of Advanced Materials Science, University of Tokyo, Kashiwa 277-8561, Japan
 \item Advanced Science Research Center, Japan Atomic Energy Agency, Tokai 319-1195, Japan
 \item Department of Physics, Tokyo Metropolitan University, Hachioji, Tokyo 192-0397, Japan
 \item Faculty of Science, University of the Ryukyus, Nishihara, Okinawa 903-0213, Japan
 \item Department of Physics, Ritsumeikan University, Kusatsu 525-8577, Japan
 \item[$^\dag$] e-mail address: {\sf shibauchi@k.u-tokyo.ac.jp}
\end{affiliations}

\vspace{-15mm}
%\begin{center}
%(Dated: \today)
%\end{center}

\newpage

\begin{abstract}
Since the 1985 discovery of the phase transition at $T_{\rm HO}=17.5$\,K in the heavy-fermion metal URu$_2$Si$_2$, neither symmetry change in the crystal structure nor magnetic ordering have been observed, which makes this ``hidden order'' enigmatic. Some high-field experiments have suggested electronic nematicity which breaks fourfold rotational symmetry, but direct evidence has been lacking for its ground state at zero magnetic field. Here we report on the observation of lattice symmetry breaking from the fourfold tetragonal to twofold orthorhombic structure by high-resolution synchrotron X-ray diffraction measurements at zero field, which pins down the space symmetry of the order. Small orthorhombic symmetry-breaking distortion sets in at $T_{\rm HO}$ with a jump, uncovering the weakly first-order nature of the hidden-order transition. This distortion is observed only in ultrapure sample, implying a highly unusual coupling nature between the electronic nematicity and underlying lattice. 
\end{abstract}

Interacting electrons in solids can generate a rich variety of phase transitions. The most essential step for elucidating the nature of a phase transition is to identify which symmetries are broken in the ordered phase. In the heavy-fermion metal URu$_2$Si$_2$ \cite{Pal85,Map86,Sch86}, tremendous efforts have been made to understand the properties of the hidden-order phase transition at $T_{\rm HO}=17.5$\,K, but its nature has been a long-standing mystery \cite{Myd11}. 
Recently, magnetic torque \cite{Oka11}, cyclotron resonance \cite{Tone_PRL,Tone_PRB}, and nuclear magnetic resonance (NMR) \cite{Kam13} measurements have suggested the occurrence of rotational symmetry breaking below $T_{\rm HO}$,  which points to an electronic nematic order with in-plane anisotropy elongated along the $[110]$ direction.  These measurements, however,  have been carried out under in-plane magnetic fields and hence we cannot rule out the possibility that the observed symmetry breaking is induced by external magnetic field.  Direct determination of the crystal symmetry in the absence of field is therefore absolutely requisite to obtain the conclusive evidence of the rotational symmetry breaking in the hidden ordered phase.   

\begin{figure*}[h]
\begin{center}
\includegraphics[width=0.5\linewidth]{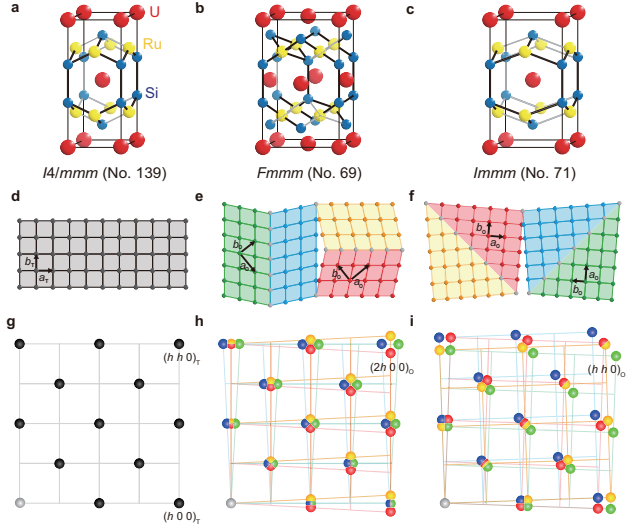}
\end{center}
\vspace{-8mm}
\caption{\small {\bf Crystals structure of URu$_2$Si$_2$ above and below the hidden-order transition.}  {\bf a}, Body-centred tetragonal $I4/mmm$ structure above $T_{\rm HO}$. {\bf b}, Orthorhombic $Fmmm$ structure revealed by the present study in the hidden-order phase below $T_{\rm HO}$. Thin solid line indicates the unit cell.  {\bf c}-{\bf f}, Schematic U atom arrangements in the basal plane ({\bf c},{\bf d}) and Bragg points in the $l=0$ plane for $h, k \geq 0$ ({\bf e},{\bf f}).  For the tetragonal $I4/mmm$ the system has a single domain ({\bf c},{\bf e}), whereas the orthorhombic $Fmmm$ structure forms four degenerate domains, which splits the Bragg $(hh0)_{\rm T}$ points into four $(2h00)_{\rm O}$ points \cite{Sch88,Tan09} ({\bf d},{\bf f}). }
\label{Fig.structure}
\end{figure*}

Owing to the electromagnetic  interaction of conduction electrons with ions of the lattice, when the electrons in a metal undergo the transition to a state which breaks one of space symmetries of the crystal, the same symmetry breaking of the underlying lattice is expected to occur.  Generally the symmetry of the low-temperature ordered phase should be lower than but belong to a subgroup of the symmetry above the transition temperature.  In the high-temperature disordered phase, URu$_2$Si$_2$ has a body-centred tetragonal crystal structure (Figs.\,\ref{Fig.structure}a,c) belonging to the $I4/mmm$ symmetry group (No.\,139 in the international tables for crystallography), which has 15 maximal non-isomorphic subgroups. Among them, the fourfold rotational symmetry is broken for two subgroups $Fmmm$ (No.\,69,  Figs.\,\ref{Fig.structure}b,d) and $Immm$ (No.\,71), which are both orthorhombic but the $ab$-plane primitive vector direction is rotated $45^\circ$ with respect to each other. These  motivate us to study $(hh0)_{\rm T}$ Bragg diffraction peaks at zero field that are most sensitive to the  $Fmmm$-type orthorhombicity (Figs.\,\ref{Fig.structure}e,f), which is compatible with the putative in-plane anisotropy elongated along the $[110]$ directions.

We use two crystals with different purities evaluated by residual resistivity ratios $RRR$; one is a crystal with $RRR \sim 10$ which is a typical value for crystals used in the previous studies \cite{Myd11} and the other is a new-generation ultraclean crystal with $RRR \sim 670$ \cite{Tone_PRL,Mat11}. Here we checked crystalline quality by X-ray at room temperature and selected ones with the sharpest Bragg peaks (Supplementary Fig.\,S1a), which ensures minimal strain effects inside the sample. Synchrotron X-ray at SPring-8 was used to analyze the crystal structure in the hidden-order phase of these crystals (see Supplementary Information). In order to achieve a very high resolution, we focus on a high-angle Bragg peak $(880)_{\rm T}$ at which our resolution is as good as $\sim 3\times10^{-5}$ (Supplementary Fig.\,S1b). Moreover, to obtain bulk information, we tune the synchrotron X-ray energy at $\sim 17.15$\,keV just below a uranium absorption edge, where X-ray attenuation length is more than 30\,$\mu$m (Supplementary Fig.\,S1c).

\section*{Results}
In Fig.\,\ref{Fig.Tdep}a, we show the temperature dependence of the $(880)_{\rm T}$ Bragg intensity as a function of the scattering vector $\bm{q}$, which is measured by the $2\theta/\theta$ mode corresponding to scans along the radial direction from the origin in the reciprocal space.  At high temperatures above $T_{\rm HO}$,  in both crystals we have a single peak with a narrow width. From the full width at half maximum, the distribution of lattice constant is estimated as $\delta a/a\sim 1.3\times 10^{-4}$, which is more than a factor of $3$ smaller than that in the previous studies \cite{Nik10,Bou11,Ker99}, indicating very high crystalline quality of our samples.
Upon entering the hidden-order phase below $T_{\rm HO}$, the data for $RRR\sim10$ sample shows no significant change in its shape, but for much cleaner sample with $RRR\sim670$ the single peak above $T_{\rm HO}$ suddenly changes to a broadened shape with clear splitting. At low temperatures the split peaks can be reasonably fitted to two Gaussian peaks with the widths comparable to the high-temperature data above $T_{\rm HO}$ (Fig.\,\ref{Fig.Tdep}b), from which the orthorhombicity 
\begin{equation}
\delta= \frac{a_{\rm O}-b_{\rm O}}{a_{\rm O}+b_{\rm O}}
=\frac{\sin\theta_2-\sin\theta_1}{\sin\theta_2+\sin\theta_1}
\end{equation}
is estimated, where $2\theta_1$ and $2\theta_2$ are the two peak angles. The change in the lattice constant $a$ for $RRR\sim10$ sample is consistent with the previous high-resolution Larmor diffraction measurements of $(400)_{\rm T}$ Bragg peak for a similar $RRR\sim10$ sample \cite{Nik10} at ambient pressure (Fig.\,\ref{Fig.Tdep}c). In sharp contrast, our new data on the ultraclean sample clearly shows a splitting into two different lattice constants $a_{\rm O}$ and $b_{\rm O}$ below $T_{\rm HO}$, evidencing the transition to the orthorhombic state. We note that these lattice constants do not track the data for the antiferromagnetic phase under pressure \cite{Nik10} and rather show an opposite trend that the averaged constant increases just below $T_{\rm HO}$. This indicates that our splitting cannot be explained by some inclusion of impurity or strain-induced phase having antiferromagnetism. We will discuss the implications of the crystal-purity dependence later, and now we focus on the data of the ultraclean sample. 

The temperature dependence of the orthorhombicity $\delta$ is demonstrated in Fig.\,\ref{Fig.Tdep}d. The orthorhombicity sets in just below $T_{\rm HO}$, indicating that the lattice symmetry change is clearly associated with the hidden-order transition. Remarkably, unlike the continuous change expected in the order parameter at a second-order phase transition, $\delta(T)$ shows a sudden jump at $T_{\rm HO}$. This strongly suggests that the transition at $T_{\rm HO}$, which has been believed to be of second order, has a first-order nature. The fact that the latent heat has not been reported \cite{Pal85,Map86,Sch86,Mat11} indicates that the first-order nature is very weak; at $T_{\rm HO}$ the discontinuity of the order parameter that characterises the low-temperature phase may be too small to be detected thermodynamically. These results lead us to conclude that the hidden order transition is a weakly first-order phase transition accompanied by lattice symmetry breaking from fourfold tetragonal to twofold orthorhombic structure. The present result is consistent with the $^{29}$Si NMR width measured under in-plane magnetic fields \cite{Kam13,Tak12}, which shows very similar temperature dependence with a clear jump at $T_{\rm HO}$ (Fig.\,\ref{Fig.Tdep}d). It has been suggested in a recent theory that the hyperfine fields at the Si site for an antiferroic $E^{-}$-type order, which has an in-plane twofold anisotropy, can lead to the NMR broadening below the transition \cite{Han12}.  Thus this naturally implies a close correspondence between the orthorhombicity and NMR broadening as found in experiments.  

To verify that the peak split originates from the lattice symmetry change to the $Fmmm$-type orthorhombic structure, we performed two-dimensional (2D) $[hk0]$ scans near the $(880)_{\rm T}$ Bragg peak (Fig.\,\ref{Fig.2D}). At 10\,K below  $T_{\rm HO}$, the data reveals a twin-peak structure (Fig.\,\ref{Fig.2D}a) in contrast to the single peak above $T_{\rm HO}$ (Fig.\,\ref{Fig.2D}b). Here the integrated intensities above and below the transition are identical within experimental error, demonstrating that the twin peaks originate from the splitting of the $(880)_{\rm T}$ Bragg peak. The elongated deformation along $[h\bar{h}0]$ direction is due to worse resolution along this line as well as finite mosaicness inevitably present in the crystal.  It should be stressed that the peak split along $[hh0]$ direction found below $T_{\rm HO}$ (Fig.\,\ref{Fig.2D}d) , which cannot be accounted for by the mosaicness, definitely indicates the appearance of two distinct lattice-plane spacings inside the crystal (reflecting the domain formation).  We also find that along $[h\bar{h}0]$ direction, although the peak does not show a clear split, it  exhibits finite broadening below $T_{\rm HO}$ (Fig.\,\ref{Fig.2D}e).  This indicates that the split occurs on two directions in the $[hk0]$ plane, indicating that the single Bragg peak above $T_{\rm HO}$ split into four peaks in the hidden-order phase. To show the consistency with the split into four peaks expected in the orthorhombic $Fmmm$-type structure \cite{Sch88,Tan09} (Fig.\,\ref{Fig.structure}f), we shift the high-temperature data at 19\,K to four directions as sketched in the inset of Fig.\,\ref{Fig.2D}c and add the four shifted data with the same weight. The calculated result shown in Fig.\,\ref{Fig.2D}c is remarkably consistent with the measured data at 10\,K (Fig.\,\ref{Fig.2D}a). The line cuts of this result also reproduce the salient features of the 10-K data (Figs.\,\ref{Fig.2D}d,e), namely the clear split along $[hh0]$ and broadened peak along $[h\bar{h}0]$. The amount of the shifts in the calculation corresponds to the orthorhombicity $\delta=6.2\times10^{-5}$, which is quantitatively consistent with the $2\theta/\theta$ scan data in Fig.\,\ref{Fig.Tdep}.  These results provide direct evidence that the lattice symmetry is lowered from the tetragonal $I4/mmm$ to orthorhombic $Fmmm$-type with the formation of four domains, and that the fourfold rotational symmetry is broken at the hidden-order transition. 

\section*{Disucssion}
\begin{figure*}[h]
\vspace{-10mm}
\begin{center}
\includegraphics[width=0.57\linewidth]{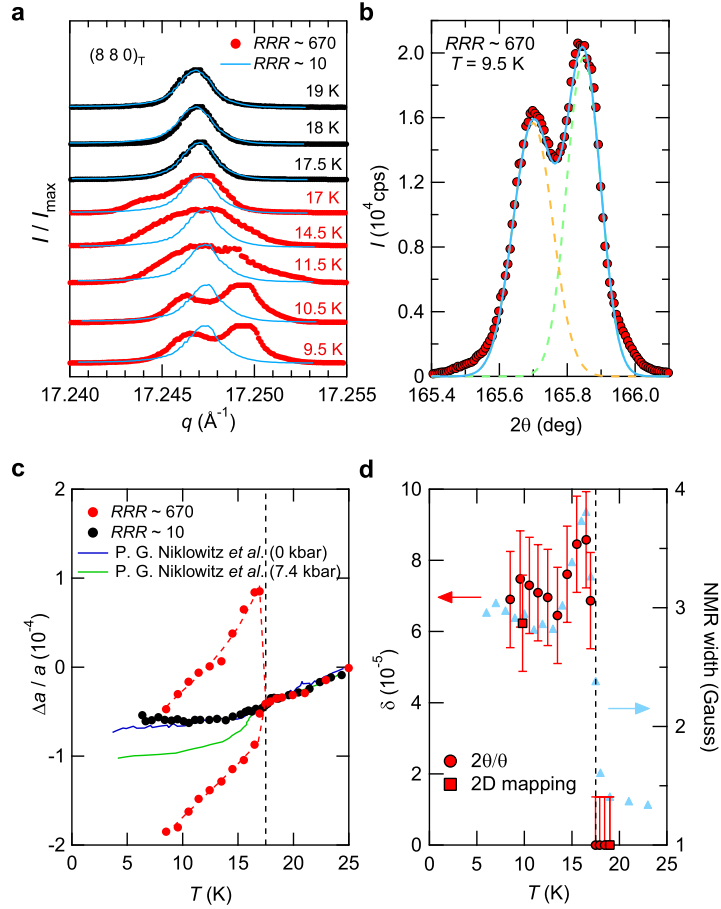}
\end{center}
\vspace{-12mm}
\caption{\small {\bf Temperature dependence of the $(880)_{\rm T}$ Bragg peak in URu$_2$Si$_2$.} {\bf a}, Intensity $I$ normalized by the peak value $I_{\rm max}$ as a function of scattering vector $q$ for two samples with different $RRR$ values (blue lines for $RRR\sim 10$, black ($T>T_{\rm HO}$) and red  ($T<T_{\rm HO}$) circles for $RRR\sim 670$). Each curve is shifted vertically for clarity. {\bf b},  The data at 9.5\,K below $T_{\rm HO}$ (circles) can be fitted to a sum (solid line) of two Gaussian functions with different lattice constants $a_{\rm O}\approx5.8290$\,\AA\, and $b_{\rm O}\approx5.8281$\,\AA\, (dashed lines).  {\bf c}, Temperature dependence of lattice constants $\Delta a(T)=a(T)-a(25\,{\rm K})$ (circles) compared with the previous report at ambient pressure (hidden order phase) and at high pressure (antiferromagnetic phase) \cite{Nik10}. Dashed lines are guides for the eyes. {\bf d}, The orthorhombicity $\delta=(a_{\rm O}-b_{\rm O})/(a_{\rm O}+b_{\rm O})$ estimated from the two-peak fitting as a function of temperature (red circles). The orthorhombicity estimated from the two-dimensional mapping at 10\,K (see Fig.\,\ref{Fig.2D}) is also plotted (red squares). The temperature dependence of the NMR line width for in-plane field (S. Kambe, private communications) is plotted for comparison (blue triangles, right axis). The dashed line marks the transition temperature $T_{\rm HO}=17.5$\,K. }
\label{Fig.Tdep}
\end{figure*}

\begin{figure*}[h]
\begin{center}
\includegraphics[width=\linewidth]{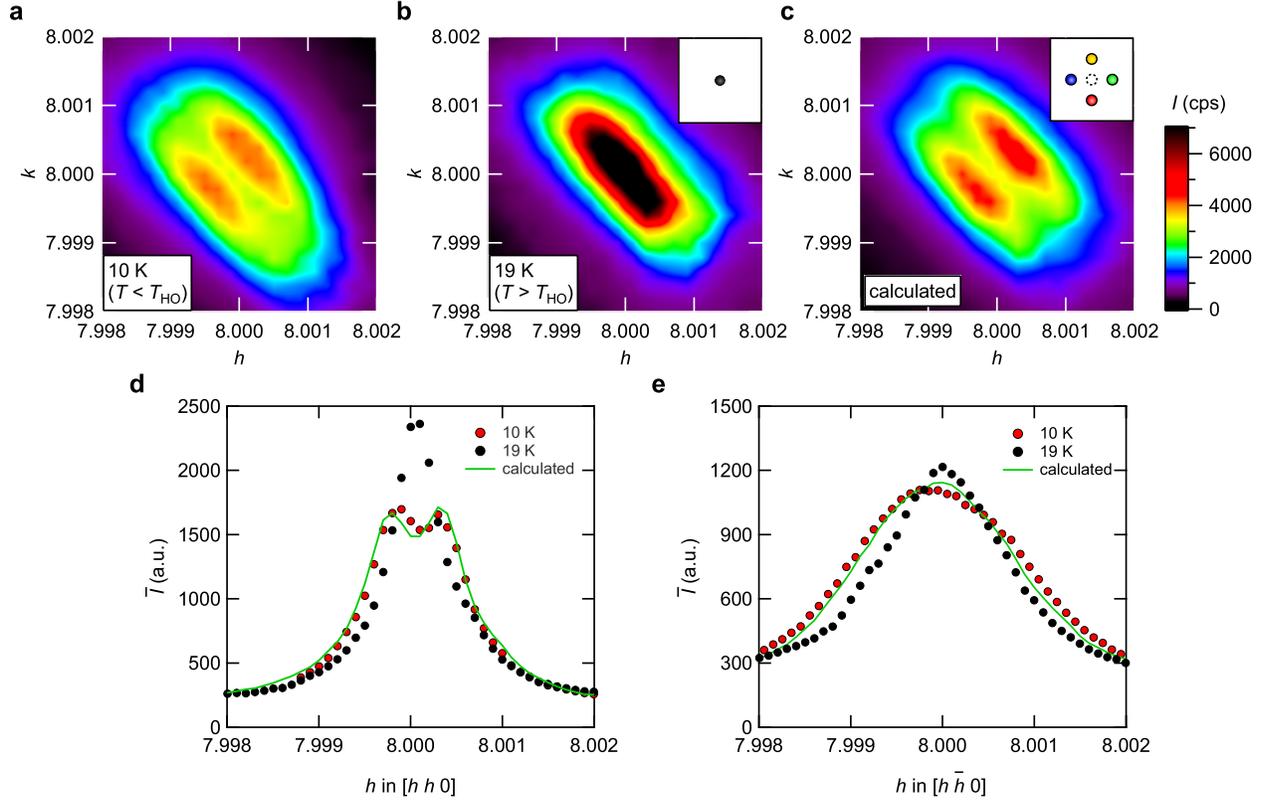}
\end{center}
\vspace{-8mm}
\caption{\small {\bf Two-dimensional mapping of the $(880)_{\rm T}$ Bragg peak.} {\bf a},  Data of $[hk0]$ scan for $7.998 \le h, k \le 8.002$ at 10\,K below $T_{\rm HO}$. {\bf b}, Data taken for the same range at 19\,K above $T_{\rm HO}$. {\bf c},  Calculated results by using the 19-K data for the orthorhombic $Fmmm$ structure with assumptions of $\delta=6.2\times10^{-5}$ and equal volumes of four domains. The inset illustrates the assumed four positions (closed circles) shifted from the original tetragonal position (dashed circle), corresponding to the four domains in the orthorhombic phase (Fig.\,\ref{Fig.structure}{\bf f}). The color bar indicates the intensity. {\bf d},  Line cuts along the $[hh0]$ direction at 10 (red circles) and 19\,K (blue circles), which are compared with the calculated one in {\bf c}  (green line). The intensity $\bar{I}$ is averaged over a constant width $\sim0.0035$\,rlu along $[h\bar{h}0]$. {\bf e}, The same plot as in {\bf d} but along the orthogonal $[h\bar{h}0]$ direction. Here the average is taken along $[hh0]$. }
\label{Fig.2D}
\end{figure*}

The in-plane electronic anisotropy elongated along the $[110]$ direction reported in the in-plane field rotation experiments \cite{Oka11,Tone_PRL,Tone_PRB,Kam13} is fully compatible with this $Fmmm$-type symmetry.  It should be emphasized that the present results are obtained at zero field, demonstrating that such electronic nematicity is not field induced. The formation of micro-domains evident from the multi-peak structure is also consistent with the above reports. 
We note that the intensity ratio of the two peaks is temperature dependent (Fig.\,\ref{Fig.Tdep}a) with the integrated intensity unchanged (Figs.\,\ref{Fig.2D}a,b). This suggests that the domain size and the position of domain walls change with temperature in the very clean crystal. This opens the possibility of `detwinning' by an external force, which may be related to the recent report that the thermal expansion anomaly at $T_{\rm HO}$ increases rapidly with application of extremely small in-plane uniaxial pressure \cite{Kam13_T}.

The present results clarify the space symmetry of the hidden-order phase, which breaks fourfold rotational symmetry. This, along with the first-order nature revealed in this study, places very tight constraints on the genuine hidden order parameter. 
Among the allowed irreducible representations for the hidden order (four non-degenerate $A_1$,  $A_2$,  $B_1$, $B_2$, and one degenerate $E$ symmetries), the orthorhombic $Fmmm$-type space group symmetry pins down that the hidden order belongs to the $E$-type, more specifically $E(\eta_a,\eta_b)$ with $\eta_a, \eta_b = \pm 1$, in which the sign of  $\eta_a\eta_b$ determines the nematic direction of the domain \cite{Tha11}. 
This establishes a solid base for the recently proposed nematic/hastatic order with in-plane anisotropy \cite{Han12,Tha11,Ike12,Cha13,Fuj11,Ris12,Rau12}.

The magnitude of orthorhombicity $\delta$ is of the order of $10^{-5}$, which is two orders of magnitude smaller than that of similar structural transitions from tetragonal $I4/mmm$ to orthorhombic $Fmmm$ phase in isomorphic BaFe$_2$As$_2$-based iron-pnictide superconductors \cite{Tan09,Kas12}. 
This smallness of the lattice change implies that the hidden-order transition is driven by an electronic ordering, and small but finite electron-lattice coupling gives rise to the lattice distortion. It should be noted that several experiments on the electronic structure provide strong evidence of the band folding over the wave vector $\bm{Q}=(001)$ \cite{Tone_PRL,Tone_PRB,Has10,
Yos10,Boa13,Men13}. Such an antiferroic ordering is expected to couple weakly to the ``ferroic'' ($\bm{Q}=0$) orthorhombic distortion. We also note that the elastic constants, which are also $\bm{Q}=0$ quantities, exhibit only small changes at $T_{\rm HO}$ \cite{Kuw97}, and it has been pointed out that these are consistent with several different symmetries including the $E(1,1)$-type state \cite{Tha11}, which is compatible with our results. 

Another remarkable finding is that the symmetry-breaking orthorhombic lattice distortion is quite sensitive to disorder (Fig.\,\ref{Fig.Tdep}a). The fact that even the low-$RRR$ samples exhibit clear signatures of the transition in the specific heat measurements \cite{Pal85,Map86,Sch86,Mat11} indicates that the $\bm{Q}=(001)$ band folding is a robust feature against disorder. However, the transition temperature $T_{\rm HO}$ shows a discernible decrease with lowering $RRR$ \cite{Mat11}, which implies that impurities can perturb the hidden order. Such unusual impurity effects may be related to the rotational degree of freedom of the nematic direction inside the $ab$ plane in the degenerate $E$-type orders. Indeed the choice of the $[110]$ direction can be made by the spin-orbit coupling, which creates small energy differences for different in-plane directions. Then impurities can induce disorder in the nematic direction, which may prevent the long-range lattice distortion through nontrivial different-$\bm{Q}$ coupling between the antiferroic order and underlying lattice. We also note that similar high sensitivity to disorder has been found in the electronic nematic phase in Sr$_3$Ru$_2$O$_7$ where the nematic anisotropy is found only in very clean samples \cite{Bor07}. Thus the impurity effects of nematic orders in strongly correlated electron systems appear to be an intriguing issue that deserves further investigations.

\begin{methods}
{\small
High-quality single crystals of URu$_2$Si$_2$ were grown by the Czochralski method in a tetra-arc furnace under argon gas atmosphere and subsequently purified by using the solid state electro-transport method under ultrahigh vacuum \cite{Mat11}. 
We used single crystals from two different batches, and the transport measurements in the crystals in these batches indicate that the residual resistivity ratios are $RRR\sim 10$ and $\sim 670$ respectively. 

The crystal structure analysis was performed by the synchrotron X-ray at SPring-8 (BL02B1). The sample was cut or crushed into small pieces and the crystalline quality of more than $\sim 30$ samples was checked at room temperature by using an imaging plate (IP). We have selected crystals with the sharpest Bragg spots at high angles for each batch. The selected best ultraclean crystal with $RRR\sim 670$ used in this study has dimensions of $\sim 70 \times 50 \times 30\,\mu$m$^3$. 
Typical Bragg peak profiles for $(880)_{\rm T}$ of this sample at room temperature are shown in Fig.\,S1a and its inset, which are taken by the four-circle diffractometer at 17.15\,keV and by the IP at 18.8\,keV, respectively. We find no tails of the peak in any direction in the IP image, and similarly circular intensity profiles were obtained for four equivalent Bragg peaks $(\pm8, \pm8, 0)$, which indicates that no significant strain is present in this crystal.

The temperature of the sample is controlled by a cryocooler equipped in the four-circle diffractometer. To expose a large portion of the crystal to the X-ray beam, we placed the sample on a fine silver wire (with a diameter of $\sim50\,\mu$m) attached to the cold head. The X-ray beam size is larger than the sample size. 
}
\end{methods}

\begin{addendum}
\item[Acknowledgements] We thank fruitful discussion with A.\,V. Balatsky, P. Chandra, P. Coleman, R. Flint, K. Ishida, S. Kambe, Y. Motome, M.-T. Suzuki, S. Takagi, P. Thalmeier, and M. Yokoyama. We especially thank S. Kambe for providing us their NMR data which we compare with our results in Fig.\,\ref{Fig.Tdep}d. 
This research was supported by KAKENHI from JSPS. The synchrotron radiation experiments were performed at BL02B1 of SPring-8 with the approval of the Japan Synchrotron Radiation Research Institute (JASRI) (Proposals No. 2011B1897, 2012A1182, and 2012B1246).
 \item[Competing Interests] The authors declare that they have no competing financial interests.
% \item[Correspondence] Correspondence and requests for materials should be addressed to T. S.~(email: shibauchi@scphys.kyoto-u.ac.jp).
\end{addendum}

\clearpage

%%%%%%%%%%%%%%%%%%Figure 1%%%%%%%%%%%%%%%%%%%%%
%%%%%%%%%%%%%%%%%%Figure 1%%%%%%%%%%%%%%%%%%%%%

%%%%%%%%%%%%%%%%%%Figure 2%%%%%%%%%%%%%%%%%%%%%
%%%%%%%%%%%%%%%%%%Figure 2%%%%%%%%%%%%%%%%%%%%%

%%%%%%%%%%%%%%%%%%Figure 3%%%%%%%%%%%%%%%%%%%%%
%%%%%%%%%%%%%%%%%%Figure 3%%%%%%%%%%%%%%%%%%%%%

\end{document}